\documentclass[12pt]{article}
\usepackage[dvips]{epsfig}
\usepackage{latexsym}
\usepackage{caption}  
\newcommand{\be}{\begin{equation}}
\newcommand{\ee}{\end{equation}}
\newcommand{\n}[1]{\label{#1}}
\newcommand{\ind}[1]{\mbox{\tiny{#1}}}
\newcommand{\ra}{\rangle}
\newcommand{\la}{\langle}
\newcommand{\para}{\parallel}

\def\bbox{{\,\lower0.9pt\vbox{\hrule \hbox{\vrule height 0.2 cm
\hskip 0.2 cm
\vrule  height 0.2 cm}\hrule}\,}}

\newcommand{\beq}{\begin{equation}}
\newcommand{\eeq}{\end{equation}}
\newcommand{\bea}{\begin{eqnarray}}
\newcommand{\eea}{\end{eqnarray}}

\begin{document}
\setlength{\unitlength}{1mm}
\title{Black Hole Radiation in the Brane World and Recoil Effect
}
\author{\\
Valeri Frolov${}^1$ and Dejan Stojkovi\'{c}${}^1$}
\maketitle

{\center
\noindent  {
$^{1}${ \em
Theoretical Physics Institute, Department of Physics, \ University of
Alberta, \\ Edmonton, Canada T6G 2J1}

}
\bigskip

\maketitle
\begin{abstract}
   A black hole attached to a brane in a higher dimensional space
emitting quanta into the bulk may leave the brane as a result of
a recoil. We study this effect. We consider black holes which
have a size much smaller than the characteristic size of extra
dimensions.  Such a black hole can be effectively described as a
massive particle with internal degrees of freedom.  We consider
an interaction of such particles with a scalar massless field
and  prove that  for a special choice of the coupling constant
describing the  transition of the particle to a state with smaller
mass the probability of massless quanta emission takes the form
identical to the probability of the black hole emission. Using
this model we calculate the probability for a black hole to leave
the brane and study its properties. The discussed recoil effect
implies that black holes which might be created in interaction of
high energy particles in colliders the thermal emission of the
formed black hole could be terminated and the energy
non-conservation can be observed in the brane experiments.
\end{abstract}

\bigskip
\vspace{3cm}

e-mails: frolov@phys.ualberta.ca, dstojkov@phys.ualberta.ca  }

\bigskip

\baselineskip=.6cm

\newpage

\section{Introduction}
\setcounter{equation}0

Recently, there has been much interest in the idea that our
$(3+1)$-dimensional universe is only a sub-manifold on which the
standard model fields are confined inside a higher dimensional
space (for a review see \cite{Rubakov}). The original ADD
(Arkani-Hamed, Dimopoulos and Dvali) idea \cite{ADD} implements
extra space as a multi-dimensional compact manifold, so that our
universe is a direct product of an ordinary $(3+1)$-dimensional
FRW (Freedman, Robertson and Walker) universe and an extra space.
This construction was primarily motivated by attractive particle
physics feature --- namely a solution to the hierarchy problem
(large difference between the Planck scale, $M_{Pl} \sim
10^{16}$TeV and the electroweak scale, $M_{EW} \sim 1$TeV). By
allowing only geometrical degrees of freedom to propagate in
extra dimensions and making the volume of the extra space large,
we can lower a fundamental quantum gravity scale, $M_*$, down to
the electroweak scale ($\sim$ TeV). The size of extra dimensional
manifold is then limited from above only by short distance gravity
experiments (current experiments do not probe any deviations from
a four-dimensional Newton's gravity law on distances smaller than
$0.2$mm). Thus, for different numbers of extra dimensions
compactified on a flat manifold (for an alternative  way of
compactification see \cite{CHM, Dienes}) the compactification
radius can vary from the fundamental length scale $M_*^{-1}$ to
the macroscopic dimensions of order $0.2$mm.

 The other option, exercised in \cite{RS}, uses a non-factorizable
geometry with a single extra dimension which can be large or even
infinite. If we are interested in solving the hierarchy problem, we
use the so-called RSI (Randal, Sundrum) scenario in which we make
extra dimension compact by introducing two branes (one with positive
and one with negative tension) with a piece of anti-de Sitter space
between them. If we put all the standard model fields on the negative
tension brane, due to exponential scaling properties of masses in
this background, we can solve the hierarchy problem by setting the
distance between the two branes  only one or two orders of magnitude
larger than the anti-de Sitter radius.

Alternatively, we can put all the standard model fields on the
positive tension brane and make the extra dimension infinite by
moving the negative tension brane to infinity (so called RSII
scenario). It is still possible to recover  four-dimensional gravity
on the positive tension brane at large distances due to a non-trivial
warp factor of the anti-de Sitter radius which sets most of the
physical volume of the extra space in a narrow region  along the
brane. This set-up does not say anything about the hierarchy problem,
but opens a possibility for new phenomena arising at the energy
scales above TeV.

A common feature of all theories with large extra dimension is
that a lot of new interesting phenomena can be expected  at the
energy scale not much above the energy currently available in
accelerators. Probably the most interesting and intriguing is the
possibility of production of mini black holes in future  collider
and cosmic rays experiments. Preliminary calculations \cite{Dim,
Giddings} indicate that the probability for creation of a mini
black hole in near future hadron colliders  such as the LHC (Large
Hadron Collider)  is so high that they can  be called  ``black
hole factories"\footnote{We should mention that there is still
some debate in literature whether or not the total probability
should be suppressed by an exponential pre-factor from
semi-classical point of view \cite{Voloshin,Soldukhin}. For the
effect we describe in this paper, this factor does not play any
significant role}.

It is straightforward to estimate the total geometrical cross
section for a production of an $(N+1)$-dimensional black hole
\cite{Dim, Giddings}. Consider two particles (partons  in the
case of the LHC) moving in opposite direction with the center of
mass energy $\sqrt{\hat{s}}$. If the impact parameter is less
than  the Schwarzschild radius of a $(N+1)$-dimensional black
hole\footnote{ \label{rot} For high energy scattering of two
particles the formation of a rotating black hole is much more
probable that the formation of a non-rotating black hole. One may
expect that mainly extremely rotating mini black holes are to be
formed in such scattering. To simplify calculations of cross
section of mini black hole production, effects connected with
rotation of the black hole are usually neglected. The same
simplification is usually made when quantum decay of mini black
holes is discussed. In the present paper we also make such an
assumption. } \be R_S= {1 \over \sqrt{\pi} M_* } \left[ { M \over
M_*} \left(  { 8 \Gamma (\frac{N}{2} ) \over N-1   }  \right)
\right]^{\frac{1}{N-2}} \ee than a black hole with a mass $M
=\sqrt{\hat{s}}$ forms. Thus, the total geometrical cross section
is \be \sigma (M) \approx \pi R_S^2 = {1 \over  M_*^2 } \left[ {
M \over M_*} \left(  { 8 \Gamma (\frac{N}{2} ) \over N-1   }
\right) \right]^{\frac{2}{N-2}} \, . \ee

Since at the LHC partons carry only a part of the total center of
mass energy in a $pp$ (proton-proton) collision, the total
production cross section is estimated as \be { d \sigma (pp
\rightarrow BH + X)  \over d M  }= {dL  \over d M} \hat{\sigma}
(ab \rightarrow BH ) |_{{\hat s}=M^2} \, . \ee Here, the
luminosity  ${dL  \over d M}$ is defined as the sum over all the
initial parton types \be {dL  \over d M}= { 2M  \over  s }
\sum_{a,b} \int_{M_{min}^2/s}^1 \frac{dx_a}{x_a}
f_a(x_a)f_b(\frac{M_{min}^2}{s x_a}) \ , \ee where $f_i(x_i)$ are
the parton distribution functions and $M_{min}$ is a minimal mass
for which this formula is applicable ($M_{min} \sim M_*$).
Numerical estimates for the total production cross section at the
LHC give for example the number of $10^7$ black holes per year if
$M_* = 1$TeV with the peak luminosity of $30 fb^{-1}$/year.

The other potential source of mini black holes are ultra high energy
cosmic rays which have been observed to interact in the Earth's
atmosphere with center of mass energy of over $100$TeV. In
particular, cosmic neutrinos could produce black hole deep in the
atmosphere, which after a rapid decay initiate quasi-horizontal
showers far above the standard model rate. Such events could be
observed in the Auger Observatory. The neutrino-nucleon scattering
cross section is calculated similarly as the $pp$ one:
\be
 \sigma (\nu N \rightarrow BH ) =
\sum_i \int_{M_{min}^2/s}^1 dx \hat{\sigma}_i(xs)f_i(x,Q) \ , \ee
where $s=2m_N E_\nu$ ($m_N$ is the nucleon mass and $E_\nu$ is the
neutrino energy), the sum goes over all partons in the nucleon,
$f_i$ are parton distribution functions and $Q$ is momentum
transfer. Calculations indicate that hundreds of such black holes
events may be observed at the Auger Observatory \cite{Feng}
before the LHC starts operating.

The differential cross-section and the total probability for a
black hole production in colliders and by  ultra high energy
cosmic rays are very well studied in literature
\cite{Dim,Giddings, Voloshin,Soldukhin,Feng, Myers,Bleicher,
Ring,Yosuke,Luis,Rizzo,Cheung,Chamblin,Kribs}.

After the black hole is formed (either at LHC or the Auger
Observatory), it decays by emitting Hawking radiation with
temperature \be T_H =  M_*  \left[ { M_{*} \over M} \left(  { N-1
\over 8 \Gamma (\frac{N}{2} )   } \right) \right]^{\frac{1}{N-2}}
{ N-2  \over 4 \sqrt{\pi} } \ . \ee

After its formation  a black hole  emits particles.
The number of particles emitted is determined by the entropy of a
black hole
\be
S={ (N-2) M \over (N-1) T_H }
\ee
For example, if $M_*=1$TeV and $N+1 =10$, a $5$TeV and $10$TeV black
holes will emit of order 30 and 50 quanta respectively.

Thermal Hawking radiation consists of two parts: (1) particles
propagating along the brane, and (2) bulk radiation. The bulk
radiation includes bulk gravitons. Usually the bulk radiation is
neglected since the total number of species which are living on
the brane is quite large (  $\sim 60$, see e.g.
\cite{Dim,Myers}). It should be noted that when the number of
extra dimensions is greater than 1 this argument may not work.
Really, the number of degrees of freedom of gravitons in the
$(N+1)$-dimensional space-time is ${\cal N}=(N+1)(N-2)/2$. For
example, for $N+1= 7$ ($3$ extra dimensions) ${\cal N}=14$.  One
may expect that if a black hole is non-rotating, emission of
particles with non-zero spin (e.g. gravitons) is suppressed with
respect to emission of scalar quanta as it happens in
$(3+1)$-dimensional space-time (\cite{Page:76}, see also
Section~10.5 \cite{FrNo} and references therein). For extremely
rotating black hole the emission of gravitons may be a dominating
effect. For example, $(3+1)$-dimensional numerical calculations
done by Don Page \cite{Page:76} (see also \cite{FrNo}) show that
the probability of emission of a graviton by an extremely
rotating black hole is about 100 times higher than the
probability of emission of a photon or neutrino. Since mini black
holes created in the high energy scattering are expected to have
high angular momentum their bulk radiation may be comparable with
(or even dominate) the radiation along the brane.

But even for small number of extra dimensions the role of bulk
graviton emission might be important. As a result of the emission
of the graviton into the bulk space, the black hole recoil can
move the black hole out of the brane.
It should be emphasized that even if the probability of the recoil
effect is not high it is of virtual importance. The reason is that
after  the black hole leaves the brane, it cannot emit brane-confined
particles anymore. Black hole radiation would be terminated and an
observer located on the brane would register  the virtual energy
non-conservation.

The aim of the present paper is to study this effect.

Don Page was first who observed that a recoil due to Hawking
radiation can be very significant  for small primordial black
holes \cite{Page}. For example, a $(3+1)$-dimensional black hole
of mass $10^{15}$g will have a recoil larger than its
Schwarzschild radius after it emits $10^{-13}$ of its mass. We
focus our attention on the recoil effect in the framework of
theories with large extra dimensions. Since the problem in its
complete scope is very complicated we make some simplifying
assumptions. First, we assume that the compactification radius of
extra dimensions in ADD scenario or anti-de Sitter radius in RS
scenario is much larger than the Schwarzschild radius of the
black hole, so that we can effectively describe the black hole by
a higher dimensional Schwarzschild solution.

Such a black hole can be effectively described as a massive particle
with internal degrees of freedom.   We characterize these different
internal states $I$ by the value of black hole mass $M_I$. Emission
of quanta of a bulk field $\varphi$ by the black hole changes its
mass and hence provides a transition $I\to J$ to the lower energy
state $J$. To simplify calculations we assume that $\varphi$ is a
bulk massless field or a set of such fields, if one wants to include
effects connected with the number of degrees of freedom of the bulk
gravitons.

In Section~\ref{2} we demonstrate that  for a special choice of
the coupling constant describing the  transition of the particle
to a state with smaller mass the probability of massless quanta
emission takes the form identical to the probability of the black
hole emission.  To describe the motion of the center of mass of
the black hole we  use $(N+1)$-dimensional wave functions.

In such a field-theoretical description of a black hole it is
possible to take into account an effect of interaction of the black
hole with a brane\footnote{Note that interaction of a black
 hole with a thin and thick test branes has been studied
 numerically in \cite{ThinDW,ThickDW} and  nucleation of
 black holes in the presence of the thick domain wall in \cite{R1,R2}.}.
 We propose a simple model for this in
Section~\ref{3}.

In Section~\ref{4} we analyze the black hole recoil effects for
models with one extra dimension. The generalization to the case of
higher number of extra dimensions is straightforward. Namely we
calculate the probability of  the black hole emission for two
cases when the black hole after emission remains on the brane and
when it leaves the brane as a result of the recoil.

Section~\ref{5} contains discussion of possible consequences of the
recoil effect.

\section{Field-Theoretical Model of an Evaporating Black Hole}

\label{2}
\setcounter{equation}0

\subsection{Probability of emission of a scalar particle with
internal degrees of freedom}

The center-of-mass motion of black hole of mass $M$ is described by a
scalar wave function $\Phi$ with an action
\be\n{2.1}
W=-{1\over 2}\, \int\, d^D x\, \left[ (\nabla \Phi)^2 +M^2\,
\Phi^2\right]\, ,
\ee
and obeying the equation
\be\n{2.2}
\Box \Phi -M^2 \Phi=0\, .
\ee
Here $D$ is the number of space-time dimensions.

We use the following
mode decomposition for the quantum field $\hat{\Phi}$
\be\n{2.3}
\hat{\Phi}(X^{A})=\int\, {d^N {\bf P}\over \sqrt{2\omega_{\bf P}}}\, {1\over
(2\pi)^{N/2}}\, \left[ e^{-i\omega_{\bf P}t+i{\bf P\,X}}\,
\hat{A}({\bf P})+ e^{i\omega_{\bf P}t-i{\bf P\,X}}\,
\hat{A}^{\dagger}({\bf P})\right]\, .
\ee
$N\equiv D-1 =3+n$ is the total number of spatial dimensions, $n$ being
the number of extra dimensions.
The bulk energy is $\omega_{\bf P}=\sqrt{{\bf P}^2+M^2}$.

Later we consider states when there is one particle in given space
volume, say $N-$cube of size $L$. Such states are more easily
described by using not continuous, but discrete levels. The
procedure is well known. By using periodic boundary conditions,
instead of waves along the $x-$axis  $\exp(ipx)/\sqrt{2\pi}$ one
has $\exp(i2\pi n/L)/\sqrt{L}$. In means that the following
modification \be\n{2.3a}
 {1\over (2\pi)^{N/2}}  \int\, d^N {\bf P}\, \to
 {1\over L^{N/2}}\, \sum_{\{ n_1,\ldots n_L\} } \,
\ee is to be applied to (\ref{2.3}).

The operators of creation and annihilation for continuous
representation (\ref{2.3}) obey the standard commutation relation
\be\n{2.5} [\hat{A}({\bf P}), \hat{A}^{\dagger}({\bf
P}')]=\delta^{N}({\bf P}-{\bf P}')\, . \ee One particle states
$|{\bf P} \ra =\hat{A}^{\dagger}({\bf P})|0\ra$ obey the
completeness condition \be\n{2.6} \int d^N {\bf P} \, |{\bf P}
\ra \, \la {\bf P} | =\hat{I}\, , \ee where $\hat{I}$ is a unit
operator.

In a similar manner we write the mode decomposition for the bulk
massless scalar field $\varphi$
\be\n{2.7}
\hat{\varphi}(X^{A})=\int\, {d^N {\bf K}\over \sqrt{2\omega}}\, {1\over
(2\pi)^{N/2}}\, \left[ e^{-i \omega t+i{\bf K\,X}}\,
\hat{a}({\bf K})+ e^{i \omega t-i{\bf K\,X}}\,
\hat{a}^{\dagger}({\bf K})\right]\, .
\ee
Here $\omega=K=|{\bf K}|$ is the bulk energy.

We choose the interaction action in the following form
\be\n{2.8}
W_{\ind{int}}=\sum_{I\ne J}\, \lambda_{IJ}\,
\int\, dX^D\, \hat{\Phi}_I(X)\, \hat{\Phi}_J(X)\, \hat{\varphi}(X)\, .
\ee
The amplitude of probability $A_{JK,I}$ of the particle (``black hole'')
transition from the initial state $I$ to the final state $J$ with
emission of a massless quantum $K$ is
\be\n{2.9}
A_{JK,I}=i\la {\bf P}_{J}, {\bf K}|W_{\ind{int}}|{\bf P}_I\ra=i\lambda_{IJ}
{2^{-3/2}\over (2\pi)^{N/2-1}}\, (\omega_{{\bf P}_I}\omega_{{\bf P}_J}
\omega)^{-1/2}\, \delta^{N}({\bf P}_I-{\bf P}_J-{\bf K})\,
\delta(\omega_{{\bf P}_I}-\omega_{{\bf P}_J}-
\omega)\, .
\ee

We assume that initially black hole is at rest, so that ${\bf
P}_I=0$ and $\omega_{{\bf P}_I}=M_I$. The probability for the
black hole to emit a quantum with energy $\omega$ per unit time
is \be\n{2.10} w(\omega)\, ={(2\pi)^N\over \Delta t\, V_N}\,
\sum_{J}\, \int\, d^N\,{\bf P}_J \int d{\bf n}_K \,\omega^{N-1}\,
|A_{JK,I}|^2\, . \ee Here $V_N$ is the space volume and $\Delta
t$ is the total time duration. The factor $ {(2\pi)^N\over V_N}$
arises because we consider the initial state when there is only
one particle in the volume $V_N$ and hence are to use the
discrete normalization for $in-$states. Since we average over
final states no additional factors will appear. As usual the
factor $\Delta t\, V_N$ is canceled by  $\delta^{N+1}(0)$ term in
$|A_{JK,I}|^2$ so that $w(\omega)$ is finite.  We also denoted
${\bf n}_K={\bf K}/K$ so that $\int d{\bf n}_K$ is the averaging
over direction of ${\bf K}$ which in our case results in the
additional factor \be\n{2.10a} \Omega_{N-1}= {2\pi^{N/2}\over
\Gamma(N/2)}\, \ee equal to the volume of $(N-1)$-dimensional
unit sphere $S^{N-1}$.

Performing integrations we get \be\n{2.11} w_I(\omega)\, =
\alpha_N\, F_I(\omega)\, , \hspace{1cm} \alpha_N={\Omega_{N-1}\,
\over 4\, (2\pi)^{N-1}}\, , \ee \be\n{2.13}
 F_I(\omega)={\omega^{N-2}\over 2 M_I}\, \sum_{J}\, \lambda_{IJ}^2\,
{\delta(M_I-\sqrt{M_J^2+\omega^2}-\omega)\over  \sqrt{M^2_J+\omega^2}}\, .
\ee

Using relation $\delta(f(x))=\delta(x-x_0)/|f'(x_0)|$ where $f(x_0)=0$,
we can rewrite (\ref{2.11}) in the form
\be\n{2.14}
 F_I(\omega)={\omega^{N-2}\over M_I}\,\sum_{J}\, \lambda_{IJ}^2\,
 \delta(M^2_J-(M^2_I-2M_I\omega))\, .
\ee
We assume that
\be\n{2.15}
M_I^2-M_J^2=\epsilon (I-J)\, .
\ee
In the limit $\epsilon \to 0$ one has continuous spectrum for black hole
mass. We assume that $\epsilon \ll T(M)/M$, where $T(M)$ is the
temperature of a black hole of mass $M$. In this case the discreteness
of the levels practically  does not affect the ``black hole'' radiation.
Since the mass $M$ is the only parameter which specify the properties
of a back hole we have
\be\n{2.16}
\lambda_{IJ}=\sqrt{\epsilon}\Lambda(M_I^2,M_J^2)\, .
\ee

We include the factor $\sqrt{\epsilon}$ into this relation to provide
the correct limit to the continuous mass spectrum case.

Changing the summation over the discrete levels $J$ by integration over
$M_J^2$ one has
\be\n{2.17}
 F_I(\omega)={\omega^{N-2}\over M_I}\,\int \, dM_J^2\, \Lambda^2(M^2_I,M^2_J)\,
 \delta(M^2_J-(M^2_I-2M_I\omega))=
 {\omega^{N-2}\over M_I}\,\Lambda^2(M^2_I,M^2_I-2M_I \omega)\, .
\ee

To summarize, the probability of emission of a massless particle
of energy $\omega$ per unit time by our ``back hole'' of mass $M$
is \be\n{2.18} p(\omega|M)={\alpha_N \omega^{N-2}\over
M}\,\Lambda^2(M^2,M^2-2M \omega)\, . \ee

\subsection{Black hole radiation}

We demonstrate now that for a special choice of the function
$\Lambda$ the probability rate (\ref{2.18}) coincides with the
probability of emission of scalar massless quanta by a black hole
of mass $M$. For this purpose let us consider $(N+1)$-dimensional
non-rotating black hole. Its metric is
\be\n{2.19} ds^2=-A\,
dt^2+{dr^2\over A}+r^2\, d\Omega_{N-1}^2\, ,
\ee
\be\n{2.20}
A=1-\left({R_0\over r}\right)^{N-2}\, . \ee The $R_0$ is the
length parameter defining the position of the horizon which is
related to the mass $M$ of the black hole as
\be\n{2.205}
M={(N-1)\Omega_{N-1}\over 16\pi G_*}\, R_0^{N-2}\, . \ee
where
$G_* = 1/M_*^{N-1} $ is a fundamental gravitational constant
determined by a fundamental energy (mass) scale $M_*$. The
surface gravity $\kappa$ is
\be\n{2.21} \kappa={1\over 2}
A'|_{R_0}={N-2\over 2R_0}\, . \ee
Thus the Hawking temperature of
the black hole is \be\n{2.22} T_H={\kappa\over 2\pi}={N-2\over
4\pi R_0}\, . \ee

Consider a scalar massless field $\varphi$ obeying the equation
\be\n{2.23} \Box \varphi =0\, . \ee Using  spherical harmonics
$Y_{lm}$ which are eigenfunctions of the Laplace operator on a
unit sphere $S^{N-1}$

\be\n{2.24} \Delta_{N-1}\, Y_{lm} = -l(l+N-2) \, Y_{lm}\, \ee one
can write the mode decomposition of $\varphi$
\be\n{2.25} \varphi
\sim \sum_{l}\, \sum_m\, {\varphi_{lm}(t,r)\over r^{n/2}}\,
Y_{lm}(\Omega)\, . \ee Here $m$ denotes a collective index which
enumerates the states with given angular momentum $l$. Modes
$\varphi_{lm}(t,r)$ obey 2D wave equations
\be\n{2.26}
({}^2\Box\varphi_{lm}-V_l)\varphi_{lm}=0, \ee
 with a potential
barrier
\be\n{2.27} V_{l}= {l(l+N-2)\over r^2}+{N-1\over 2}\left[
{N-3\over 2}\,{A\over r^2} +{A'\over r}\right] \, . \ee
The
operator ${}^2\Box$ is the box-operator in the 2D metric of
$(t,r)$ sector.

Let $r_*$ be the tortoise like radial coordinate \be\n{2.28}
r_*=\int {dr\over A}\, , \ee and $\varphi_{lm}(t,r)=\exp(-i\omega
t)\, \varphi_{\omega,lm}(r)$ be a monochromatic wave solution of
(\ref{2.26}) then $\varphi_{\omega,lm}(r)$ obeys the equation

\be\n{2.29} \left[{d^2\over dr_*^2}+[\omega^2 -A\,V_{l}]\right]
\varphi_{\omega,lm}=0\, . \ee
This equation can be rewritten in
the dimensionless form by using variables $x=r/R_0$, \
$x_*=r_*/R_0$,\ $\varpi=R_0 \omega$ as follows
\be\n{2.30}
\left[{d^2\over dx_*^2}+[\varpi^2 -W_{l}(x)]\right]
\varphi_{\omega,lm}=0\, , \ee
where
\be\n{2.31} W_{l}(x)=
(1-{1\over x^{N-2}})\left[{A\over x^2}+\left({N-1\over
2}\right)^2\, {1\over x^{N}}\right]\, .
\ee
\be\n{2.32}
A=l(l+N-2)+{(N-1)(N-3)\over 4}\, . \ee

Denote by $T_l(\varpi)$ the  amplitude probability to penetrate
the potential barrier $W_{l}$ for a mode with frequency $\omega$.
Then the greybody factor is defined as
\be\n{2.33}
\Gamma(\varpi)=\sum_{l=0}^{\infty}\, \Gamma_l(\varpi) \,
,\hspace{1cm} \Gamma_l(\varpi)=\sum_m |T_l(\varpi)|^2\, .
\ee

The probability of emission of the massless field quanta of energy
$\omega$ by the black hole is (see e.g. \cite{FrNo}, section
10.4.4)
\be\n{2.34} P(\omega|M)={1\over 2\pi}
\sum_{l=0}^{\infty}\, \sum_m\, {(e^{\varpi\beta} -1)\,
|T_l(\varpi)|^2\over (e^{\varpi\beta} -1 +|T_l(\varpi)|^2)^2}\, .
\ee
Here $\beta$ is the dimensionless inverse temperature
\be\n{2.35} \beta=(TR_0)^{-1}=4\pi/(N-2)\, .
\ee

\begin{figure}
\begin{tabular}{cc}
\epsfig{file=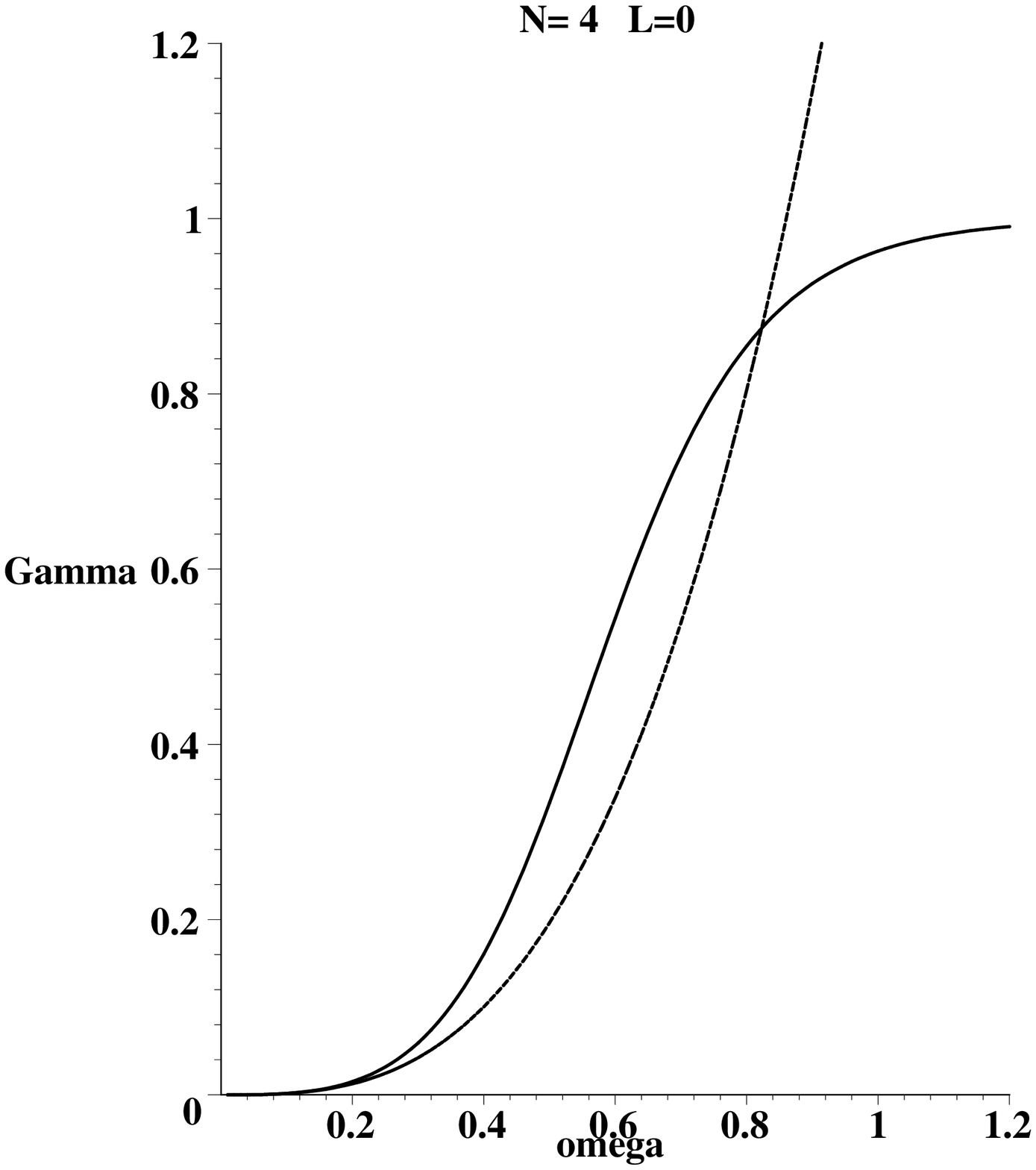, width=4.5cm} \hspace{2cm}&\hspace{2cm}
\epsfig{file=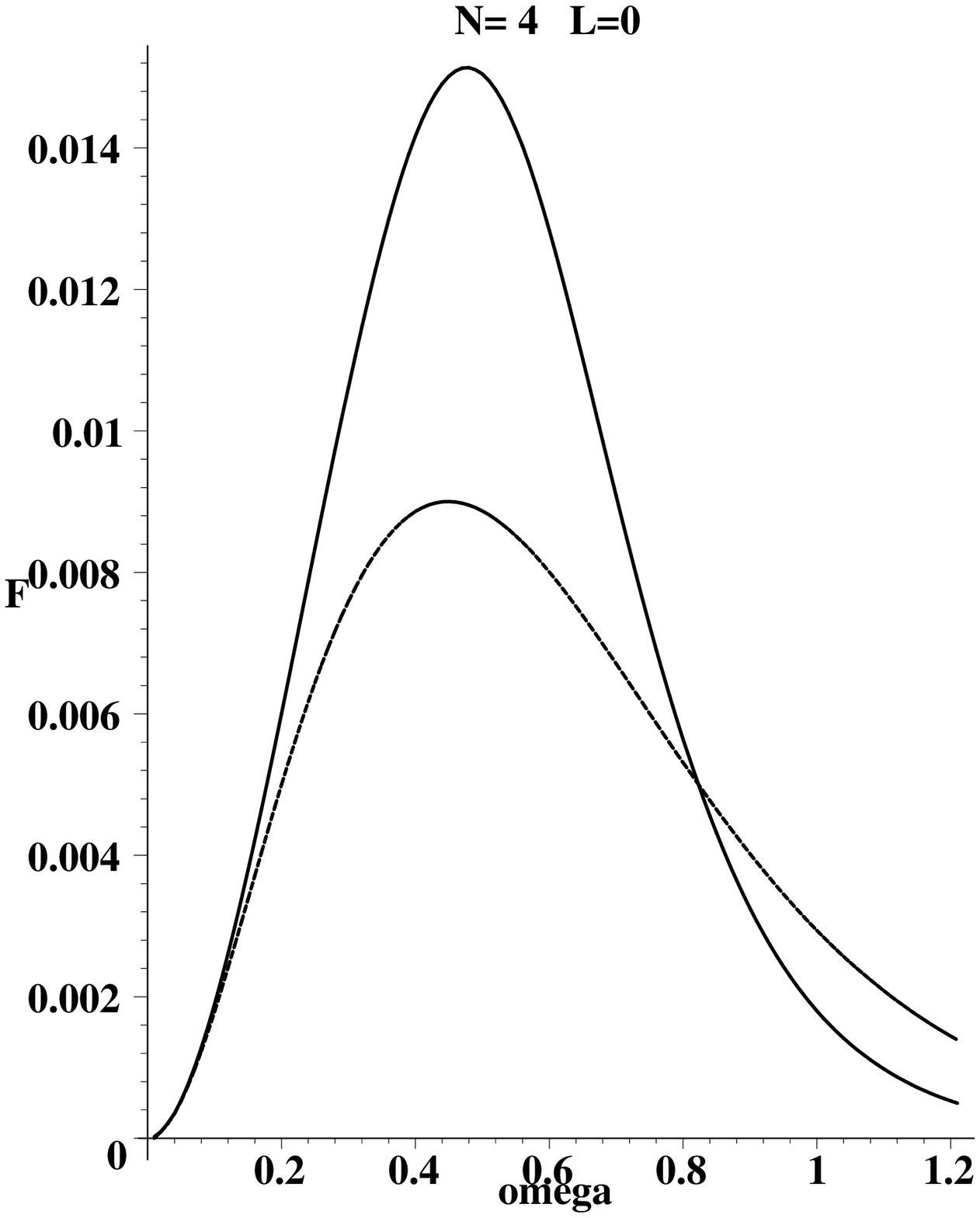, width=4.5cm}\\
{\bf a1} \hspace{2cm}&\hspace{2cm} {\bf b1}\\
\epsfig{file=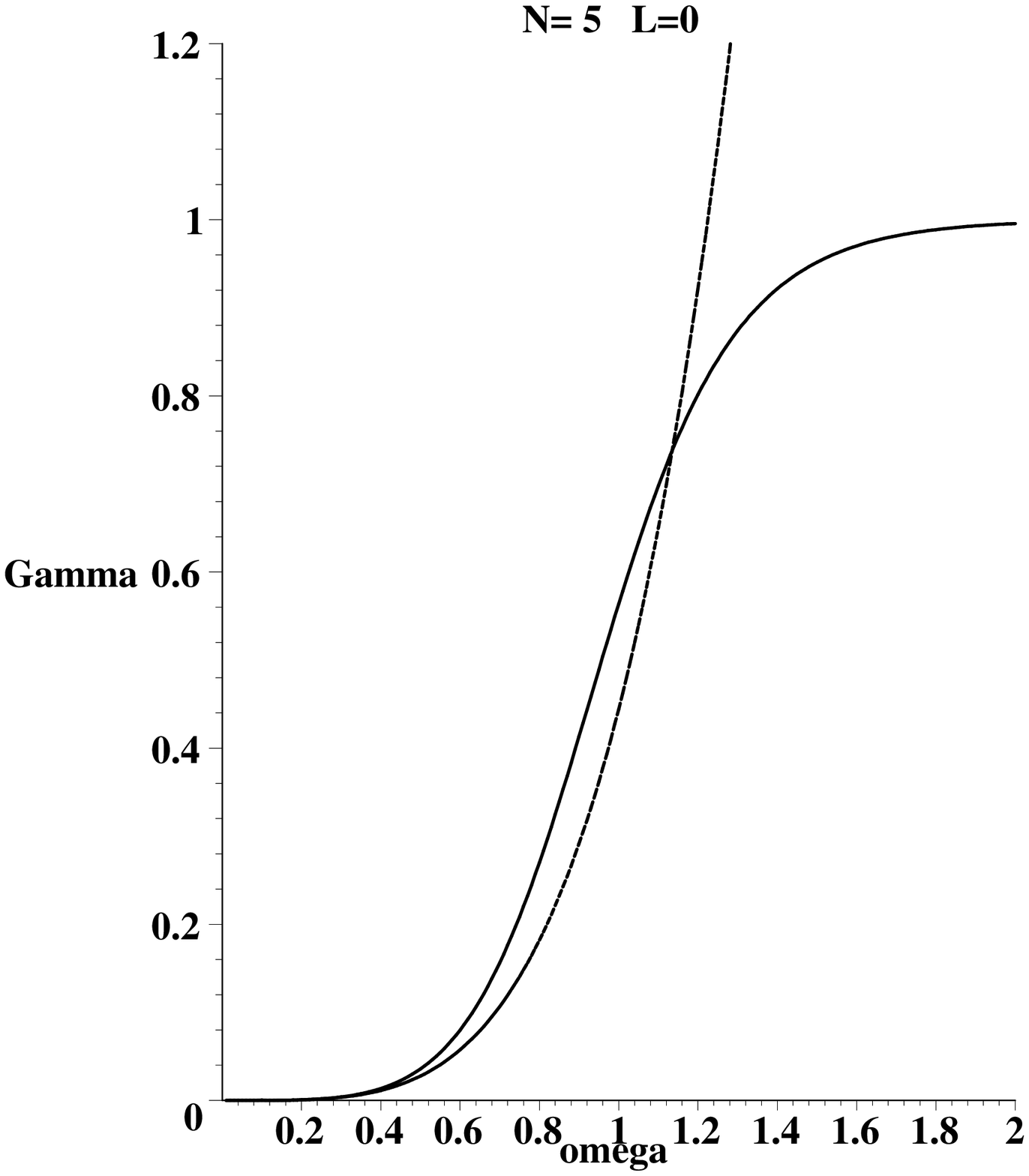, width=4.5cm} \hspace{2cm}&\hspace{2cm}
\epsfig{file=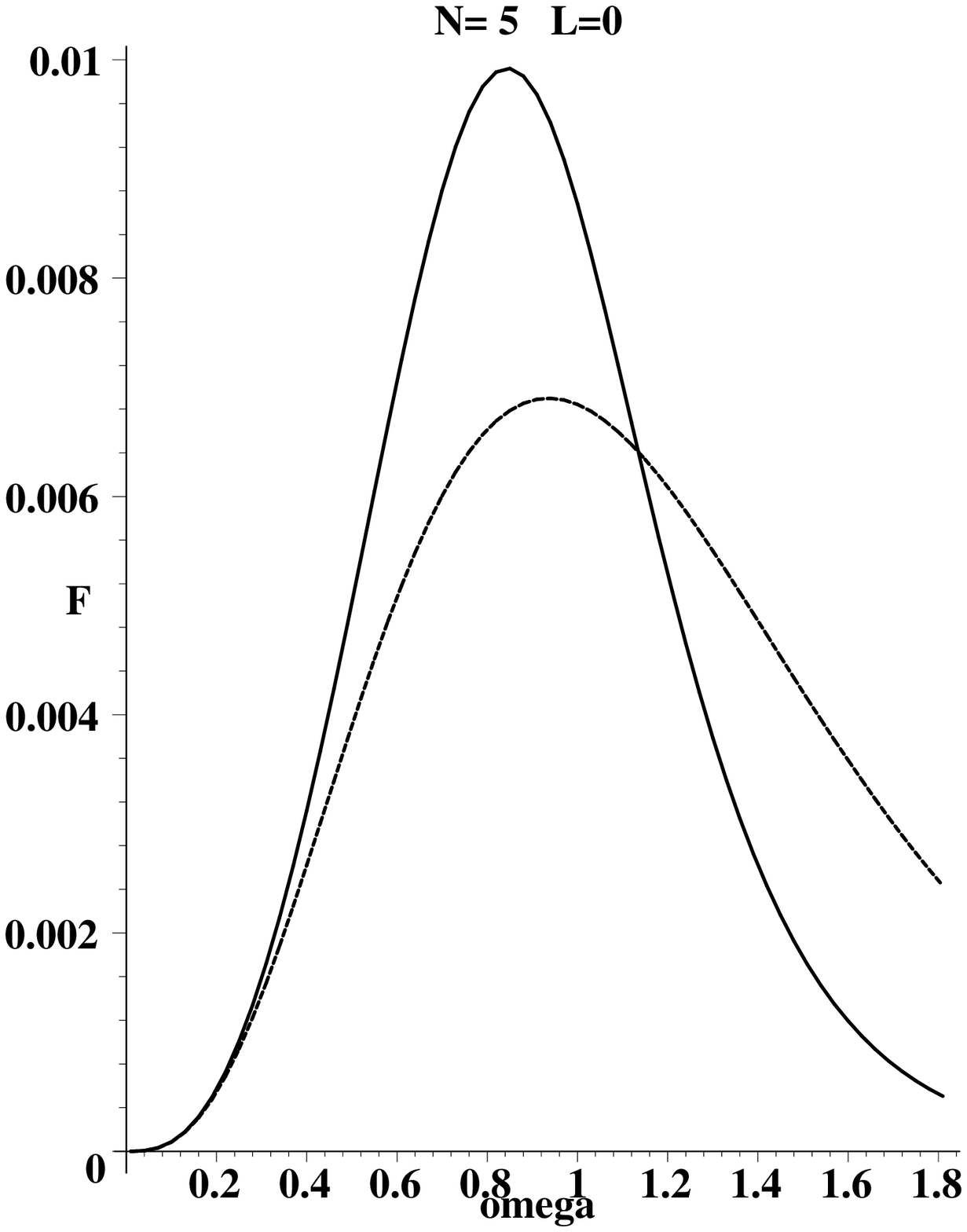, width=4.5cm}\\
{\bf a2} \hspace{2cm}&\hspace{2cm} {\bf b2}\\
\epsfig{file=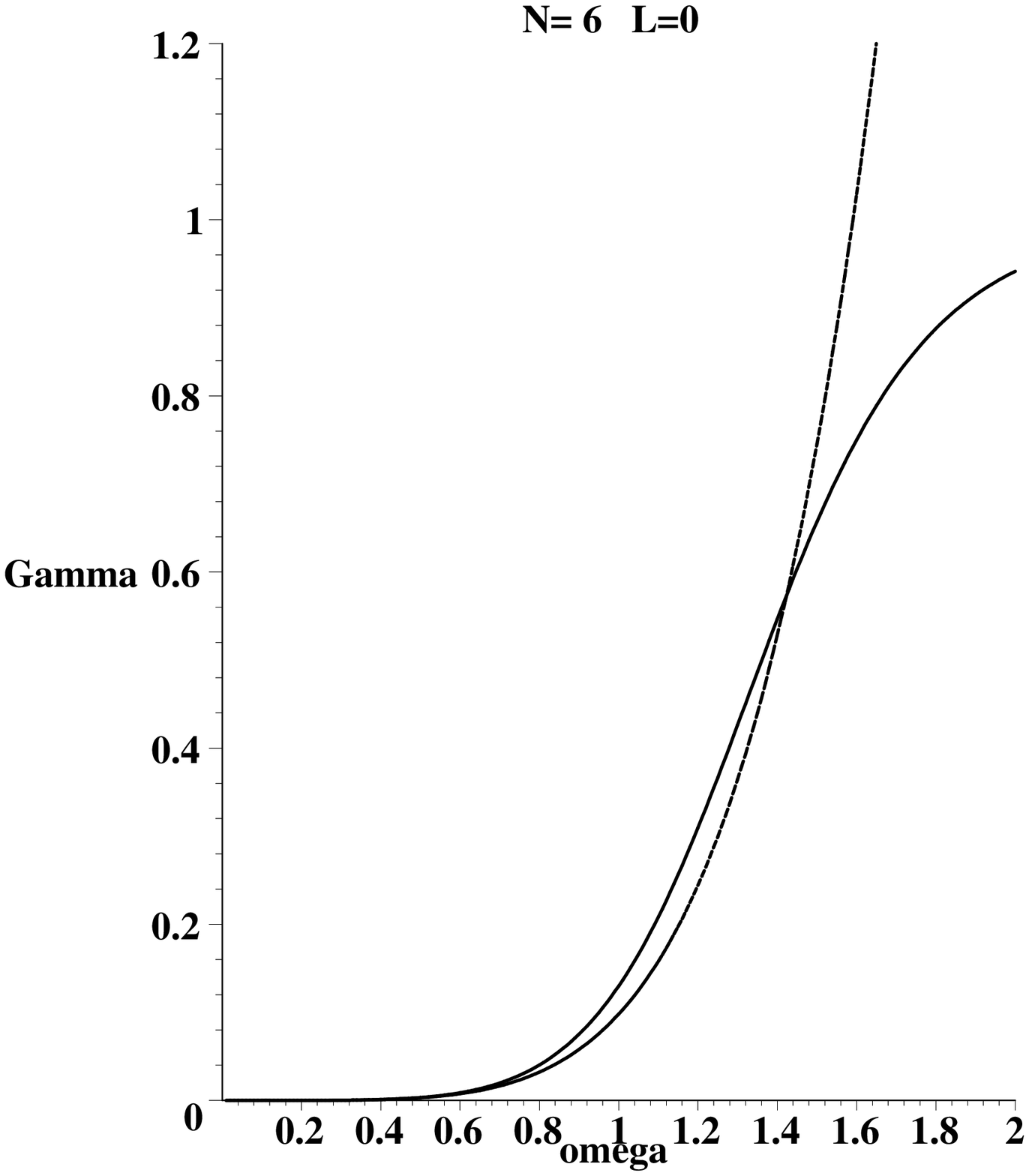, width=4.5cm} \hspace{2cm}&\hspace{2cm}
\epsfig{file=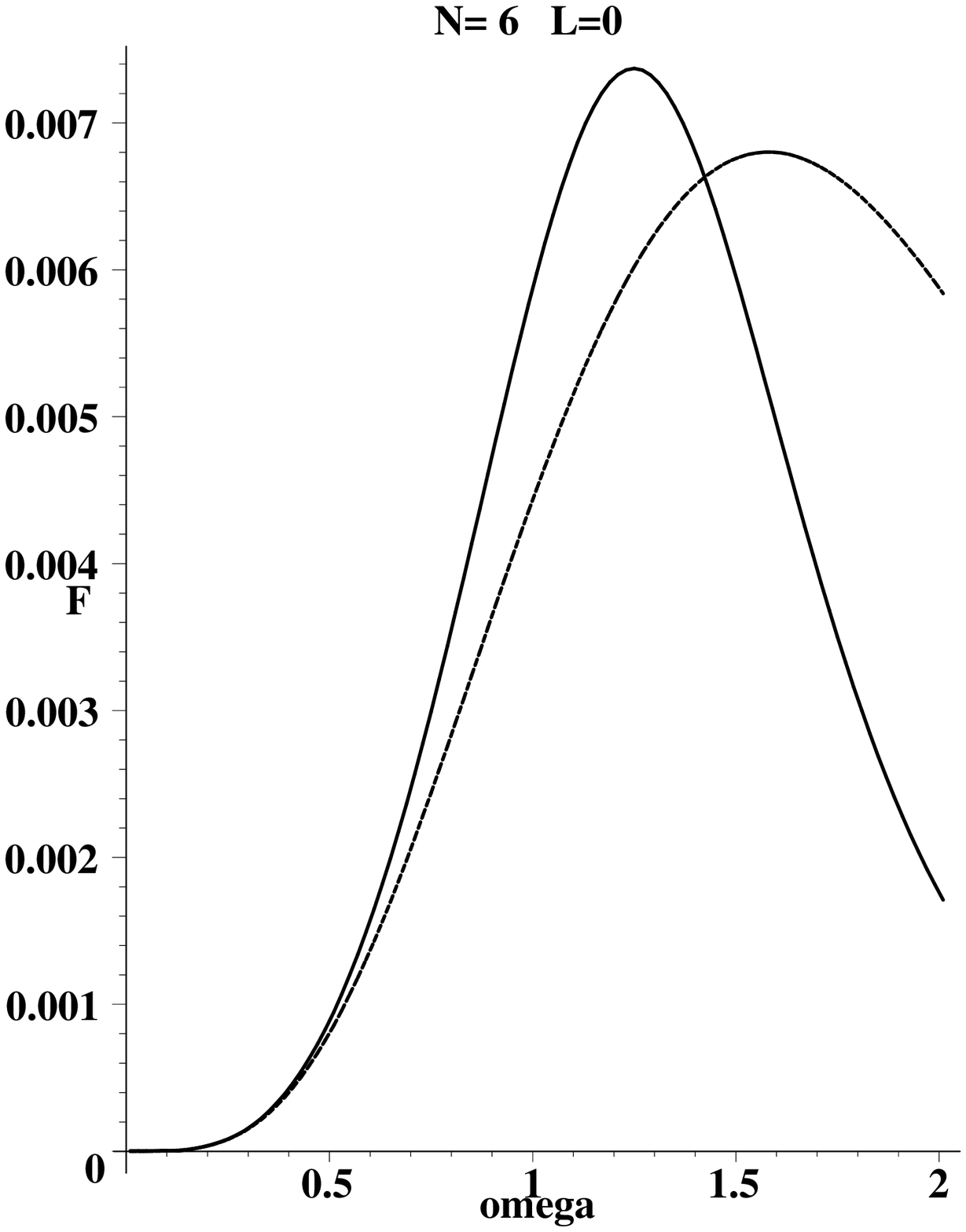, width=4.5cm}\\
{\bf a3} \hspace{2cm}&\hspace{2cm} {\bf b3}
\end{tabular}
\caption{Greybody factor $\Gamma$ as a function of
$\varpi=R_0\omega$ for $N=4$ (a1), $N=5$ (a2), and $N=6$ (a3).
The curves representing $\Gamma$ in the vicinity of $\omega=0$
are well approximated by the curves shown at the same figures. At
large values of $\omega$  \, $\Gamma\to 1$. The figures b1, b2,
and b3 show the function ${\cal F}_(\varpi)$ for $N=4$, $N=5$,
and $N=6$ respectively. Similar plots for the low energy
approximation are also shown.} \label{curves}
\end{figure}

$s$-modes (that is modes with $L=0$) give the main contribution to
the Hawking radiation. For this reason we shall keep only $L=0$
contribution in (\ref{2.34}). Moreover, numerical calculations show
that $|T_0(\varpi)|^2 \ll e^{\varpi\beta} -1$. For $N=4,5,6$  the functions
\be\n{2.36}
{\cal F}_(\varpi)={\Gamma_0(\varpi)\over e^{\varpi\beta} -1}\,
\ee
are shown at Fig.~\ref{curves}.

Under these conditions one has
\be\n{2.37}
P(\omega|M)={1\over 2\pi}\, {\cal F}_(\varpi)\, .
\ee

For small frequencies one has (see e.g. \cite{GuKlTs:97,KaMa:02} and
references therein)
\be\n{2.38}
\Gamma_{\ind{lf}}(\varpi)={\pi\varpi^{N-2}\over 2^{N-3}\Gamma^2(N/2)}\, .
\ee
Plots presented at Fig.~\ref{curves} demonstrate that the probability
function ${\cal F}(\varphi)$ can be reasonably well approximated if one use
$\Gamma_{\ind{lf}}(\varpi)$ instead of the exact value of the grey
body factor. In what follows we shall use this {\em low frequency
approximation}.

Comparing (\ref{2.18}) with (\ref{2.37}), in the low frequency
approximation, one can conclude that if we want a decaying
massive particle $M$ to emit massless quanta with the same
probability as a evaporating black hole one must choose
\be\n{2.39} \Lambda^2(M^2,M^2-2M\omega)= {\gamma_N\, M\,
R_0^{N-2}\over e^{\beta\varpi}-1}\, , \ee where \be\n{2.39a}
\gamma_N={(2\pi)^{N-1}\over \pi^{N/2}\, 2^{N-3}\, \Gamma(N/2)}\,
. \ee Denote \be\n{2.40} \xi=M^2\, ,\hspace{1cm}
\zeta=M^2-2M\omega\, , \ee then we can write \be\n{2.41}
\Lambda^2(\xi, \zeta)={\gamma_N\, f(\xi)\over{\displaystyle \exp
\left[{\beta R_0(\xi)(\xi-\zeta)\over 2\sqrt{\xi}}\right]-1}}\, ,
\ee where \be\n{2.42} f(\xi)=M\, R_0^{N-2}\, , \hspace{1cm}
R_0(\xi)=\left[{16\pi G_* \over
(N-1)\Omega_{N-1}}\right]^{1/(N-2)}\, \xi^{1/(2(N-2))}\, . \ee

In the next section we shall use the expression for
$\Lambda^2(\xi,\zeta)$ for the case $N=4$. Calculations give
\be\n{2.44} \Lambda^2(\xi,\zeta)={2^5 \over 3}\, {G_* \, \xi \over
{\displaystyle \exp \left[ b\, \sqrt{G_*}\, (\xi - \zeta)
/\xi^{1/4}\right]-1}}\, , \ee where $b=\sqrt{8\pi/3}$.

\section{Interaction of a black hole with a brane}

\label{3}
\setcounter{equation}0

Suppose  now that there exists a brane  representing our physical
world embedded in higher dimensional universe. For simplicity we
assume that the brane has a co-dimension 1, i.e.  there exist
only one extra dimension. In different models it is usual to
consider either compactified extra dimensions (see e.g
\cite{ADD}) or to assume that the space out of the brane is
anti-de Sitter one \cite{RS}). We assume here that the size $R_0$
of a $(N+1)$-dimensional black hole is much smaller than the
compactification scale or the anti-de Sitter curvature radius. In
this case  one can neglect the action of the environment onto the
metric of the black hole. Certainly, this is true only for the
gravitational field near the black hole. At far distances, the
gravitational field of the black hole is modified.

When a black hole is close to the brane or intersect the latter,
its gravitational field is modified. Unfortunately till now there
is no exact solutions for a black hole on the (3+1)-brane (see
however \cite{EmHoMy} where such a solution was found for
(2+1)-brane). In what follows we study the recoil effect caused
by the emission of massless field quanta in the bulk space. As
earlier, we neglect the effects connected with the spin and
consider emission of the scalar massless field. Moreover we use a
simplified model to take into account the effects connected with
the interaction of the black hole with the brane. Namely, we
neglect the effects of the brane gravitational field on the
Hawking radiation, and use one parameter, similar to a chemical
potential, to describe the interaction of the black hole with the
brane. Below we describe this model.

As earlier we consider the black hole as a point particle with
internal degrees of freedom like in the section~2.1 but now we rewrite
its action (\ref{2.1}) as follows
\be\n{3.1}
W=-{1\over 2}\, \int\, d^D x\, \left[ (\nabla \Phi)^2 +U\,
\Phi^2\right]\, ,
\ee
\be\n{3.2}
U=M^2-2\mu\, \delta(y)\, .
\ee
The field $\Phi$  obeys the equation
\be\n{3.3}
\Box \Phi -U \Phi=0\, .
\ee

To construct the mode decomposition of the field operator
$\hat{\Phi}$ we first consider the following eigenvalue problem
\be\n{3.5}
\left[{d^2\over dy^2}+2\mu\, \delta(y)\right]\, \Phi = \lambda\,
\varphi\, .
\ee
Denote a solution of this equation in the interval of negative and
positive values of $y$ as $\Phi^{(-)}(y)$ and $\Phi^{(+)}(y)$,
respectively. Then the junction conditions at $y=0$ are
\be\n{3.6}
\Phi_{>}(0)=\Phi_{<}(0)\, ,\hspace{1cm}
\Phi_{>}'(0)-\Phi_{<}'(0)=-2\mu\,\Phi_{>}(0) \, .
\ee

It is easy to see that there is only one level with positive
$\lambda=\mu^2$. A wave function of the corresponding bound state
is

\be\n{3.7} \Phi^{(0)}(y)=\sqrt{ \mu } \, e^{-\mu\, |y|} \, . \ee
We choose the coefficient so that $\Phi_0(y)$ obeys the following
normalization condition \be\n{3.8} \int_{-\infty}^{\infty}\, dy\,
|\Phi_0(y)|^2=1\, . \ee

This normalization condition is in agreement with the
normalization of fields in the previous section.

For negative $\lambda$ the spectrum is continuous. We denote
$\lambda=-p_{\perp}^2$. Solving the scattering problem for the
$\delta$-like potential one obtains the following set of solutions
\be\n{3.9}
\Phi^{(+)}_{p_{\perp}}(y)=
\left\{ \begin{array}{ll}
e^{ i p_{\perp} y}-{\mu\over \mu +ip_{\perp}}\,e^{ -i p_{\perp} y} \ ,
& y<0\ , \\ \\
{ip_{\perp}\over \mu +ip_{\perp}}\,e^{ i p_{\perp} y} \ , &y>0 \ ,
\end{array}
\right.
\ee
\be\n{3.10}
\Phi^{(-)}_{p_{\perp}}(y)=
\left\{ \begin{array}{ll}
{ip_{\perp}\over \mu +ip_{\perp}}\,e^{ -i p_{\perp} y} \  ,
& y<0\ , \\ \\
e^{ -i p_{\perp} y}-{\mu\over \mu +ip_{\perp}}\,e^{ i p_{\perp} y} \
 &y>0 \ .
\end{array}
\right.
\ee
obeying the normalization conditions
\be\n{3.11}
{1 \over 2 \pi } \int_{-\infty}^{\infty}\, dy\, \Phi^{(+)}_{p_{\perp}}(y)\,
\Phi^{(+)*}_{p_{\perp}'}(y)=
{1 \over 2 \pi } \int_{-\infty}^{\infty}\, dy\, \Phi^{(-)}_{p_{\perp}}(y)\,
\Phi^{(-)*}_{p_{\perp}'}(y)=\delta(p_{\perp}-p_{\perp}')\, .
\ee
The other similar integrals vanish.

The complete set of modes consists of two types of solutions. The
first type are solutions describing a black hole attached to the
brane which can freely propagate only along it. These solutions
are \be\n{3.12} \Phi^{(0)}_{{\bf
p}_{\para}}({X})={e^{-i\tilde{\omega}t}\over
\sqrt{2\tilde{\omega}}  \, (2\pi)^{(N-1)/2}}\, \Phi_0(y)\,
e^{i{\bf p}_{\para}{\bf x}_{\para}}\, , \ee where \be\n{3.13}
\tilde{\omega}^2=\tilde{M}^2+p_{\para}^2\, ,\hspace{1cm}
\tilde{M}^2=M^2-\mu^2\, . \ee The second type of modes are bulk
modes of the form \be\n{3.14} \Phi^{(\pm)}_{{\bf
P}}({X})={e^{-i\omega t}\over \sqrt{2\omega}\, (2\pi)^{N/2}}\,
\Phi^{(\pm)}_{p_{\perp}}(y)\, e^{i{\bf p}_{\para}{\bf
x}_{\para}}\, , \ee \be\n{3.15}
\omega^2=M^2+p_{\para}^2+p_{\perp}^2\, . \ee Here and later we
use the following notations. $P^{A}$ ($A=1,...,N$) is total bulk
momentum which has components $p^i_{\para}$ ($i=1,2,3$) along the
brane, and $p^a_{\perp}$ ($a=1, ... ,n$) in the bulk direction.
We also denote $x^i$ coordinated along the brane, and $y^a$ bulk
coordinates. Thus we have \be\n{2.4} {\bf P\, X}={\bf
p}_{\para}\, {\bf x}+\bf{p}_{\perp}\, {\bf y}\, . \ee

The field operator decomposition takes the form
\[
\hat{\Phi}(X)=\int\, d^{N-1}p_{\para}\, \left[
\Phi^{(0)}_{{\bf p}_{\para}}({X})\, \hat{B}({\bf p}_{\para})+
\bar{\Phi}^{(0)}_{{\bf p}_{\para}}({X})\, \hat{B}^{\dagger}({\bf
p}_{\para})\right]
\]
\be\n{3.16}
{ }\quad \quad +
\sum_{\pm}\, \int\, d^N {\bf P}\, \left[
\Phi^{(\pm)}_{{\bf P}}({X})\, \hat{A}_{\pm}({\bf P})+
\bar{\Phi}^{(\pm)}_{{\bf P}}({X})\, \hat{A}_{\pm}^{\dagger}({\bf P})
  \right]\, .
\ee
For the scalar massless field we shall use the decomposition
(\ref{2.7}). This means that we neglect possible interaction of scalar
quanta with the brane.

It should be emphasized that in a general case the constant $\mu$
may depend on $M$. The form of this dependence can be calculated when
one obtains a solution describing a black-hole-brane system.
An important example of a 3-dimensional black hole attached to 2-brane
was studied in \cite{EmHoMy}.
Unfortunately at the moment  no generalization of these solutions to
higher dimensions are known. We shall discuss different options of the
choice of $\mu(M)$ in the Section~\ref{5}.

\section{Radiation of a black hole attached to the brane and the
recoil effect}

\label{4}
\setcounter{equation}0

\subsection{Matrix elements and  probability for leaving the brane}

As earlier the amplitude of probability $A_{JK,I}$ of the particle
(``black hole'') transition from the initial state $I$ to the final
state $J$ with emission of a massless quantum $K$ is
\be\n{3.17}
A_{JK,I}^{off}=i\la {\bf P}_{J}, {\bf K}|W_{\ind{int}}|{\bf P}_I\ra \, ,
\ee
where $W_{\ind{int}}$ is given by (\ref{2.8}). We choose as the
initial state $|I\ra$ a state of the black hole at rest at the brane
\be\n{3.18}
|I\ra =\hat{B}^{\dagger}({\bf p}_{\para}=0)\,|0\ra \, .
\ee
We are interested in those final states of a black hole when as a
result of the recoil it leaves the brane and moves in the bulk space.
Thus we choose
\be\n{3.19}
|J,K\ra =\hat{A}^{\dagger}_{\pm}({\bf P})\,\hat{a}^{\dagger}({\bf K})|0\ra \, .
\ee

Simple calculations give \be\n{3.20}
A_{JK,I\pm}^{off}=i\lambda_{IJ} {2^{-3/2}\over
(2\pi)^{(N-1)/2}}\, (\tilde{M}_I \,\omega_{{\bf P}_J}
\omega)^{-1/2}\, \delta^{N-1}({{\bf p}_{\para}}_J+{\bf
k}_{\para})\, \delta(\tilde{M}_I-\omega_{{\bf P}_J}- \omega_k)
C_{\pm}\, , \ee where \be\n{3.21} C_{\pm}= \int\, dy \,
\Phi^{(0)}(y)\, e^{ik_{\perp}y}\,
\Phi^{(\pm)}_{p_{J,\perp}}(y)\, . \ee
Calculating the integral in
(\ref{3.21}) we get
\be\n{3.22} C_{+}={\sqrt{\mu}\over \mu
+ip_{J,\perp}}\, \left[ {\mu +ip_{J,\perp}\over \mu
+ip_{J,\perp}+ik_{\perp}}- {\mu \over \mu
-ip_{J,\perp}+ik_{\perp}}+ {ip_{J,\perp}\over \mu
-ip_{J,\perp}-ik_{\perp}}\, \right]\, , \ee
\be\n{3.225}
C_{-}={\sqrt{\mu}\over \mu +ip_{J,\perp}}\, \left[ {\mu
+ip_{J,\perp}\over \mu +ip_{J,\perp}-ik_{\perp}}- {\mu \over \mu
-ip_{J,\perp}-ik_{\perp}}+ {ip_{J,\perp}\over \mu
-ip_{J,\perp}+ik_{\perp}}\, \right]\, . \ee

The probability per unit time of the black hole emission which
results in the recoil taking the black hole away from the brane is
\be\n{3.23} w_{off}\, ={(2\pi)^{N-1}\over \Delta t
V_{N-1}}\sum_{J}\, \int\, d^N\,{\bf P}_J \int d{\bf K} \,
|A_{JK,I}^{off}|^2\, . \ee By taking the integral over ${\bf
p}_{\para}$  one gets \be\n{3.24} w_{off}={\Omega_{N-2}\over 8\,
 (2\pi)^{N}\, \tilde{M}_I}\, \sum_J\, |\lambda_{IJ}|^2\,
\int_0^{\infty} dp_{\perp}\, \int_{-\infty}^{\infty} dk_{\perp}\,
\int_0^{\infty} dk_{\para}k_{\para}^{N-2}\, {\delta(\tilde{M}_I-
\omega_p\, -\omega_k\, )\over \omega_p\, \omega_k\,}\, |C|^2\, .
\ee Here \be\n{3.25}
\omega_p=\sqrt{M_J^2+k_{\para}^2+p_{\perp}^2}\, ,\hspace{1cm}
\omega_k=\sqrt{k_{\para}^2+k_{\perp}^2}\, , \ee \be\n{3.26}
|C|^2=|C_+|^2+|C_-|^2={32\,\mu^3\, k_{\perp}^2\, p_{\perp}^2\,
(\mu^2+k_{\perp}^2+p_{\perp}^2)\over
(\mu^2+k_{\perp}^2+2k_{\perp}p_{\perp}+p_{\perp}^2)^2\,
(\mu^2+k_{\perp}^2-2k_{\perp}p_{\perp}+p_{\perp}^2)^2\,
(\mu^2+p_{\perp}^2)}\, . \ee

We use the relation (see Section~2.1)
 \be\n{3.27} {1\over 2}\,
\sum_J\, |\lambda_{IJ}|^2\,{\delta(\tilde{M}_I- \omega_p\,
-\omega_k\, )\over \omega_p}\, =\Lambda^2(\tilde{M}_I^2, m^2)\, ,
\ee
where
\be\n{3.28}
m^2=\tilde{M}_I^2-2\tilde{M}_I\sqrt{k_{\para}^2+k_{\perp}^2}+k_{\perp}^2-p_{\perp}^2\,
. \ee
We omit the subscript ``I", i.e. $\tilde{M}_I \equiv
\tilde{M}$.

Rewrite now (\ref{3.24}) as follows \be\n{3.29}
w_{off}={\Omega_{N-2}\over 4\, (2\pi)^{N}\, \tilde{M}}\,
\int_0^{\infty} dp_{\perp}\, \int_{-\infty}^{\infty} dk_{\perp}\,
\int_0^{\infty} dk_{\para}{k_{\para}^{N-2}\over \omega_k}\,
\Lambda^2(\tilde{M}^2,m^2)\, |C|^2\, .
 \ee

For $N=4$ (that is for the (3+1)-brane in 5-dimensional
space-time) after using (\ref{2.44}) we have \be\n{3.30}
w_{off}={2\over 3\pi^3}\,G_* \tilde{M}\, \int_0^{\infty}
dp_{\perp}\, \int_{-\infty}^{\infty} dk_{\perp}\, \int_0^{\infty}
{dk_{\para}\,k_{\para}^{2}\over \omega_k}\,\, |C|^2 \, Q\, , \ee
\be\n{3.31} Q={1 \over \exp\left[b\sqrt{G_* /{\tilde M}}
(2\tilde{M}\sqrt{k_{\para}^2+k_{\perp}^2}+
k_{\perp}^2-p_{\perp}^2)  \right] -1}\, , \ee where
$b=\sqrt{8\pi/3}$.

One can expect that the main contribution to $w_{off}$ is given
by low frequency bulk modes since the higher frequency ones are
suppressed by the thermal factor. According to this, we neglect
$k_{\perp}^2$ and $p_{\perp}^2$ terms in the exponent with
respect to $\tilde{M}k$ term and approximate $Q$ by the expression
\be\n{3.32} Q_0(k)={1\over \exp\left[(2b\sqrt{G_* \tilde{M}}\,
k)  \right] -1}\, , \ee where $k=\sqrt{k_{\para}^2+k_{\perp}^2}$.
After this simplification the integral over $p_{\perp}$ in
(\ref{3.30}) can be easily taken \be\n{3.33} \int_0^{\infty}
dp_{\perp}\,|C|^2 = {2\pi\,k_{\perp}^2(8\mu^2+k_{\perp}^2)\,
\over \, (k_{\perp}^2+4\mu^2)^2}\, . \ee We substitute this
expression into (\ref{3.30}) and using the fact that the integrand
is an even function of  $k_{\para}$ extend the region of
integration over this variable to $(-\infty, \infty)$. Then
passing to the polar coordinates $k_{\para}=k\sin\phi$,
$k_{\perp}=k\cos\phi$ we get \be\n{3.34} w_{off}={2\over 3 \pi^2
}\, \tilde{M} \, \int_0^{\infty}\, dk\, k^4\, Q_0(k)\, Z(k)\, ,
\ee where \be\n{3.35} Z(k)=\int_0^{2\pi}\, d\phi\, {\sin^2\phi\,
\cos^2\phi\, (k^2\, \cos^2\phi+8\mu^2)\over (k^2\, \cos^2\phi
+4\mu^2)^2}\, . \ee The integral (\ref{3.35}) can be taken
exactly \be\n{3.36} Z(k)={\pi\,(\sqrt{k^2+4\mu^2}-2\mu)\over k^2\,
\sqrt{k^2+4\mu^2}}\, . \ee Thus we obtain the following
representation for $w_{off}$ \be\n{3.37} w_{off}={2 \over 3 \pi
}\, G_*  \tilde{M}\, \int_0^{\infty}\, dk\,
{k^2\,(\sqrt{k^2+4\mu^2}-2\mu)\over \sqrt{k^2+4\mu^2}} {1 \over
\exp(a\, k)-1}\, . \ee where $a=2b \sqrt{G_* \tilde{M}}= 4
\sqrt{{2\pi \over 3}} \sqrt{G_* \tilde{M}}$.

Finally, we introduce dimensionless variables \be\n{3.38} \chi =a
k \, ,\hspace{1cm} \nu=2 \mu a = 8 \sqrt{{2\pi \over 3}} \mu
\sqrt{G_* \tilde{M}}\, , \ee and write $w_{off}$ in the form
\be\n{3.39} w_{off}(\nu) ={\mu \over 8 \pi^2 \nu }
\int_0^{\infty}\, d\chi\, {\chi^2\,(\sqrt{\chi^2+\nu^2}-\nu)\over
\sqrt{\chi^2+\nu^2}} {1\over \exp(\chi)-1 }\, . \ee

\subsection{Matrix elements and  probability of remaining  on the brane}

Calculation of the probability that the black hole remains  on the brane
is analogous to the calculation  given in the previous section.
The amplitude of probability $A_{JK,I}$ for the particle
(``black hole'') to stay on the brane after emitting
a massless quantum $K$ is:

\be\n{4.1}
A_{JK,I}^{on}=i\la {\bf P}_{J}, {\bf K}|W_{\ind{int}}|{\bf P}_I\ra \, ,
\ee
where $W_{\ind{int}}$ is again  given by (\ref{2.8}). We choose as the
initial state $|I\ra$ a state of the black hole at rest at the brane
\be\n{4.2}
|I\ra =\hat{B}^{\dagger}({\bf p}_{\para}=0)\,|0\ra \, .
\ee
We are interested in those final states of a black hole when as a
result of the recoil it remains on the brane, i.e. does not move
to the bulk space.
Thus we choose
\be\n{4.3}
|J,K\ra = \hat{B}^{\dagger}({\bf p}_{\para} \neq 0)  \,\hat{a}^{\dagger}({\bf K})|0\ra \, .
\ee

Repeating  calculations from previous section we get: \be\n{4.4}
A_{JK,I\pm}^{on}=i\lambda_{IJ} {2^{-3/2}\over  (2\pi)^{N/2-1}}\,
(\tilde{M}_I \,\omega_{{\bf P}_J} \omega)^{-1/2}\,
\delta^{N-1}({{\bf p}_{\para}}_J+{\bf k}_{\para})\,
\delta(\tilde{M}_I-\omega_{{\bf P}_J}- \omega_k) D(k_{\perp})\, ,
\ee with \be\n{4.5} D(k_{\perp})= \int\, dy \, \Phi^{(0)}(y)\,
e^{ik_{\perp}y}\,  \Phi^{(0)}(y)\, . \ee Simple integration gives
\be\n{4.6} D(k_{\perp})= \frac{4 \mu^2}{4 \mu^2 + k_{\perp}^2} \,
. \ee

The probability per  unit time of the black hole emission which
does not take the black hole away from the brane is \be\n{4.7}
w_{on}\, ={(2\pi)^{N-1}\over \Delta t V_{N-1}}\sum_{J}\, \int\,
d^N\,{\bf P}_J \int d{\bf K} \, |A_{JK,I}^{on}|^2\, . \ee
Integrating over ${\bf p}_{\para}$  we have

\be\n{4.8} w_{on}={\Omega_{N-2}\over 8\,   (2\pi)^{N-1}\,
\tilde{M}_I}\, \sum_J\, |\lambda_{IJ}|^2\, \int_{-\infty}^{\infty}
dk_{\perp}\, \int_0^{\infty} dk_{\para}k_{\para}^{N-2}\,
{\delta(\tilde{M}_I- \omega_p\, -\omega_k\, )\over \omega_p\,
\omega_k\,}\, |D(k_{\perp})|^2\, . \ee with\be\n{4.9}
\omega_p=\sqrt{M_J^2+k_{\para}^2}\, ,\hspace{1cm}
\omega_k=\sqrt{k_{\para}^2+k_{\perp}^2} \, ,\hspace{1cm}
p_{\perp} =0 \, . \ee

As before, we use the relation
\be\n{4.10}
{1\over 2}\,  \sum_J\, |\lambda_{IJ}|^2\,{\delta(\tilde{M}-
\omega_p\,
-\omega_k\, )\over \omega_p}\, =\Lambda^2(\tilde{M}^2, m^2)\, ,
\ee
where
\be\n{4.11}
m^2=\tilde{M}_I^2-2\tilde{M}_I\sqrt{k_{\para}^2+k_{\perp}^2}+k_{\perp}^2 \,
.\ee
We omit the subscript ``I", i.e. $\tilde{M}_I
\equiv \tilde{M}$.

Now, (\ref{4.7}) becomes \be\n{4.12} w_{on}={\Omega_{N-2}\over
4\,   (2\pi)^{N-1}\, \tilde{M}}\, \int_{-\infty}^{\infty}
dk_{\perp}\, \int_0^{\infty} dk_{\para}{k_{\para}^{N-2}\over
\omega_k}\, \Lambda^2(\tilde{M}^2,m^2)\, |D(k_{\perp}) |^2\, . \ee

For $N=4$ (that is for the (3+1)-brane in 5-dimensional
space-time), using (\ref{2.44}) we have
\be\n{4.13} w_{on}={4\over
3 \pi^2}\,G_* \tilde{M}\, \int_{-\infty}^{\infty} dk_{\perp}\,
\int_0^{\infty} {dk_{\para}\,k_{\para}^{2}\over
\omega_k}\,\, |D(k_{\perp}) |^2\, Q\, , \ee
\be\n{4.14} Q={1\over
\exp\left[b\sqrt{G_* /{\tilde M}}
(2\tilde{M}\sqrt{k_{\para}^2+k_{\perp}^2}+ k_{\perp}^2)  \right]
-1}\, , \ee
 where $b=\sqrt{8\pi/3}$.

Neglecting $k_{\perp}^2$ term
 in the exponent with respect to $\tilde{M}k$ term
we approximate $Q$ by the expression \be\n{4.15} Q_0(k)={1\over
\exp\left[(2b\sqrt{G_* \tilde{M}}\, k)  \right] -1}\, . \ee where
$k=\sqrt{k_{\para}^2+k_{\perp}^2}$. Passing to the polar
coordinates $k_{\para}=k\sin\phi$, $k_{\perp}=k\cos\phi$ we get
\be\n{4.16} w_{on}={2\over 3 \pi^2}\, \tilde{M} \,
\int_0^{\infty}\, dk\, k^2\, Q_0(k)\, Z(k)\, , \ee where
\be\n{4.17} Z(k)=\int_0^{2\pi}\, d\phi\, {16 \mu^4 \sin^2\phi\,
\over (k^2\, \cos^2\phi +4\mu^2)^2}\, . \ee The integral
(\ref{4.17}) can be taken exactly \be\n{4.18} Z(k)={2\pi \mu
\over \sqrt{k^2+4\mu^2}}\, . \ee We obtain the final
representation for $w_{on}$ as \be\n{4.19} w_{on}={4 \over 3
\pi}\, G_*  \tilde{M}\, \int_0^{\infty}\, dk\, {k^2\, \mu \over
\sqrt{k^2+4\mu^2}} {1 \over \exp(a\, k)-1}\, . \ee

where $a=2b \sqrt{G_* \tilde{M}}= 4 \sqrt{{2\pi \over 3}}
\sqrt{G_* \tilde{M}}$. Introducing  dimensionless variables
\be\n{4.20} \chi =a k \, ,\hspace{1cm} \nu=2 \mu a = 8 \sqrt{ {2
\pi \over 3}} \mu \sqrt{G_* \tilde{M}} \, , \ee we write $w_{on}$
in the form \be\n{4.21} w_{on}(\nu)= {8 \mu \over \pi}
\int_0^{\infty}\, d\chi\, {\chi^2\, \over \sqrt{\chi^2+\nu^2}}
{1\over \exp(\chi)-1 }\, . \ee

\begin{figure}
\centerline{
\epsfig{file=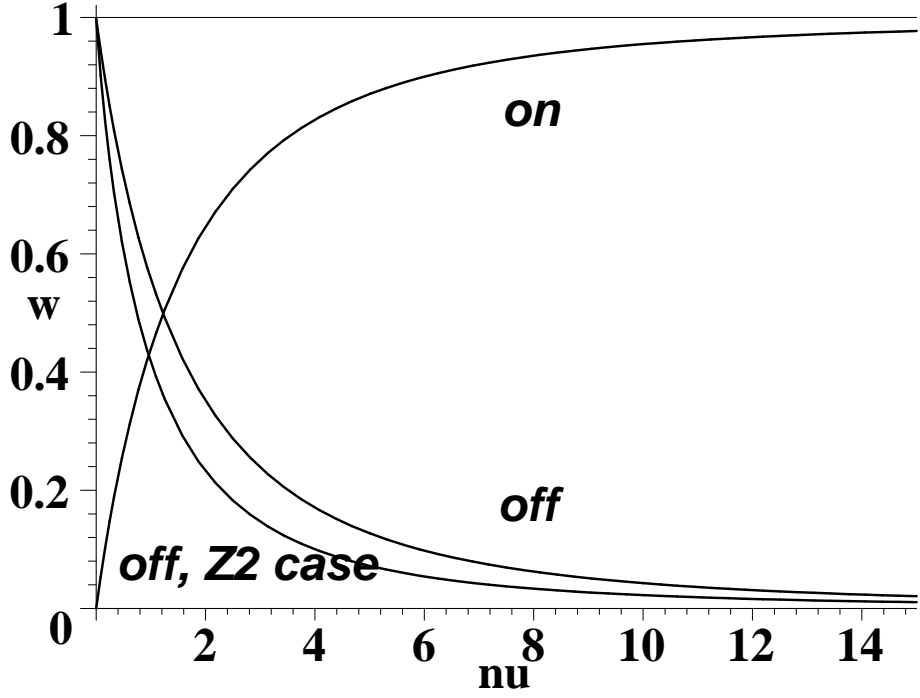, width=12cm}                    }
\caption{Functions $w_{off}(\nu)/w$ and $w_{on}(\nu)/w$.}
\label{w}
\end{figure}

It is easy to see that the sum \be\n{4.22}
w=w_{off}(\nu)+w_{on}(\nu)\, \ee does not depend on $\nu$. In
fact $w$ coincides with the total probability of  emission of  a
massless field quantum by the 5-dimensional black hole:
\be\n{4.23} w=\int_0^{\infty}\, d\omega \, P(\omega|M)\, , \ee
where $P(\omega|M)$ is given by (\ref{2.37}). The plots of the
functions $w_{off}(\nu)/w$ and $w_{on}(\nu)/w$ are shown at
Fig.~\ref{w}. Analysis shows that, for large $\nu$ (or $\mu$),
the probability $w_{off}$ falls off as $1/\nu$, while for small
$\nu$ (or $\mu$) the (normalized) probability $w_{off}$ goes to
$1$.

\subsection{Brane world with $Z_2$ symmetry}

Here, we repeat the calculations of previous sections for the
case when the forms of wave functions are restricted by
additional $Z_2$ symmetry, like in RS scenarios.  By imposing the
symmetry under transformation $y \rightarrow -y$, we restrict the
wave function of the continuous spectrum (\ref{3.9}) and
(\ref{3.10})  to a symmetric linear combination

\be\n{4.24}
\Phi^{sym}_{p_{\perp}}(y)=\Phi^{(+)}_{p_{\perp}}(y)+
\Phi^{(-)}_{p_{\perp}}(y) =
\left( e^{- i p_{\perp} |y|}-{\mu - i p_{\perp}\over \mu +ip_{\perp}}\,
e^{ i p_{\perp} |y|} \right) \ .
\ee
Also, the bulk part of the wave function of the emitted massless scalar
particle must attain a symmetric form, i.e. $ e^{ i k_{\perp} |y|}$.

First, we recalculate the probability for leaving the brane.
Since main steps are the same as earlier, we give here only the final
result for probability to leave the brane

\be\n{4.26} w^{sym}_{off}={4 \over 3 \pi}\, G_*  \tilde{M}\,
\int_0^{\infty}\, dk\, \left( -4 \mu\, \sqrt{k^2+4 \mu^2}  +8
\mu^2 + k^2 \right) {1 \over \exp(a\, k)-1}\, . \ee

where $a=2b \sqrt{G_* \tilde{M}}= 4 \sqrt{{2\pi \over 3}}
\sqrt{G_* \tilde{M}}$. In  dimensionless variables \be\n{4.27}
\chi =a k \, ,\hspace{1cm} \nu=2 \mu a = 8 \sqrt{{2\pi \over 3}}
\mu \sqrt{G_* \tilde{M}}\, , \ee this is \be\n{4.28}
w^{sym}_{off} (\nu) ={1 \over 4 \pi  } \int_0^{\infty}\, d\chi\,
\left( -\nu \sqrt{\chi^2+\nu^2} + \nu^2 + {1 \over 2}
\chi^2               \right) {1\over \exp(\chi)-1 }\, . \ee

Similarly, for the probability to remain on the brane we get

\be\n{4.29} w^{sym}_{on}={16 \over 3 \pi}\,  G_*
\tilde{M}\, \int_0^{\infty}\, dk\, \left(  \mu \sqrt{k^2+4 \mu^2}  -2 \mu^2
\right) {1 \over \exp(a\, k)-1}\, . \ee
where $a=2b \sqrt{G_*
\tilde{M}}= 4 \sqrt{{2\pi \over 3}} \sqrt{G_* \tilde{M}}$.
In dimensionless variables (\ref{4.27}) this is \be\n{4.31}
w^{sym}_{on}(\nu) ={1\over 4 \pi  } \int_0^{\infty}\, d\chi\,
\left( \nu \sqrt{\chi^2+\nu^2}  - \nu^2 \right) {1\over \exp(\chi)-1
}\, . \ee

Once again the sum
\be\n{4.32}
w=w_{off}(\nu)+w_{on}(\nu)\,
\ee
does not depend on $\nu$ and coincides with the total
probability of emission of a massless field quantum by the
5-dimensional black hole.

\section{Discussion}
\label{5}
\setcounter{equation}0

In literature, the process of evaporation of mini black holes in
brane world models is well studied. The pattern of such a black
hole decay is markedly different from any other standard model
event. There are even very detailed calculations estimating the
energy spectrum and ratio between emitted particles (leptons,
fotons, hadrons...)\cite{Giddings} . It is also claimed that,
because of very little missing energy, the determination of the
mass and the temperature of the black hole may lead to a test of
Hawking's radiation. The recoil effect discussed above may change
some of these predictions.  Certainly the most important
observable effect of a black hole recoil is a suddenly disrupted
evaporation and local non-conservation of energy.

We developed a phenomenological model for description of the
recoil effect. This model contains an important parameter $\mu$
which play the role of the chemical potential. In the general
case, $\mu$ depends on the mass of a black hole $M$ and the
tension $\sigma$ of the brane, $\mu=\mu(M,\sigma)$.
Unfortunately, to determine this dependence is not an easy task.
One can expect that in the limit $\sigma\to 0$, that is for a
test brane, $\mu\to 0$. In this case, it is very likely that the
black hole leaves the brane as soon as emits first quanta with
non-zero bulk momentum.

We can also estimate $\mu$ for small $\sigma$ as follows.
Consider two different states. First, a brane with a black hole
of radius $R_0$, and the second, when the same black hole is out
of the brane. The second configuration has extra energy $\Delta E
\sim \sigma R_0^3$ (for 1 extra dimension). One can identify
$\Delta E$ with $M-\tilde{M}$ (compare with (\ref{3.12})). This
gives \be \label{est}\mu^2\approx 2M\Delta E\, , \ee or \be
\mu\sim \sigma^{1/2}\, R_0^{5/2}\sim \sigma^{1/2}\, M_0^{5/4}\, .
\ee

In the other limit of infinitely heavy brane, or a brane with
$Z_2$ symmetry, the process of a black hole leaving the brane
reminds to a black hole splitting into two symmetric black holes
in the ``mirror'' space. Classically this process is forbidden in
a higher dimensional space-time for the same reason as in
$(3+1)$-dimensional space-time in connection with non-decreasing
property of the entropy. In the presence of cosmological
constant, such an effect may become possible as a tunneling
process. These arguments show that in this case $\mu \rightarrow
\infty$ or is exponentially large, and the recoil effect is
suppressed. This feature could help in distinguishing between the
two different scenarios of extra dimensions --- ADD and RS. Also,
if the general conclusion that the recoil effect may be important
for branes of small tension is correct, it opens an interesting
possibility of using the experiments with decay of mini black
holes to put restrictions on the brane tension.

The above order of magnitude estimates of the parameter $\mu$ do
not follow from the exact calculations and therefore might not be
very reliable\footnote{Our attention was drown by Don Page who
pointed out that more detailed analysis shows that the estimate
(\ref{est}) is quite valid in the case of the brane of a
co-dimension $3$ and higher (three or more extra spatial
dimensions). For co-dimensions $1$ and $2$ result might depend on
a regime in which we extract the black hole, i.e. whether it is an
adiabatic process or not.}. In any case further investigation is
required, which will be the topic of eventual further
publications.

We note that the results from section \ref{2}, about the field
theoretical model describing a black hole as a point radiator, are
quite general and are not restricted to brane world models.

\bigskip

\vspace{12pt} {\bf Acknowledgments}:\ \  The authors are grateful
to Don Page and Glenn Starkman for stimulating discussions. This work was partly
supported  by  the Natural Sciences and Engineering Research
Council of Canada. The authors are grateful to the Killam Trust
for its financial support.

\end{document}